# Synergistic effect in two-phase laser procedure for production of silver nanoparticles colloids applicable in ophthalmology

A. S. Nikolov (1)*, N. E. Stankova (1), D. B. Karashanova (2), N. N. Nedyalkov (1), E. L. Pavlov (1), K. Tz. Koev (1, 3), Hr. Najdenski (4), V. Kussovski (4), L. A. Avramov (1), C. Ristoscu (5), M. Badiceanu (5, 6), and I. N. Mihailescu (5)

(1) Institute of Electronics, Bulgarian Academy of Sciences, 72 Tsarigradsko Chaussee, Sofia 1784, Bulgaria;
(2) Institute of Optical Materials and Technologies, Bulgarian Academy of Sciences, G. Bonchev Street, bl. 109, Sofia 1113, Bulgaria;
(3) Department of Ophthalmology, Medical University - Sofia, 8 Bjalo more, Sofia 1000, Bulgaria;
(4) The Stefan Angeloff Institute of Microbiology, Bulgarian Academy of Sciences, 26 Georgi Bonchev str., 1113 Sofia, Bulgaria;
(5) National Institute for Lasers, Plasma and Radiation Physics, PO Box MG-36, RO-77125, Magurele, Ilfov, Romania;
(6) University of Bucharest, Faculty of Physics, RO-77125, Magurele, Ilfov, Romania

* Corresponding author
E-mail address: anastas_nikolov@abv.bg

**Abstract**

This work reports on the production of Ag nanoparticles (AgNPs) in water solution based upon two-phase pulsed laser procedure for ophthalmological therapeutic approaches. In this case, the AgNPs should be ≤ 10 nm and have a narrow size distribution. Nanoparticles of this sized-scale are capable to penetrate the complex ocular barriers, ensuring effective non-invasive drug delivery to retina. Moreover, the ocular irritation, which is currently associated to the conventional ocular drug administration routes, would be avoided.

In the first phase, AgNPs larger than 20 nm were fabricated via laser ablation of a Ag target under water by irradiation with a fundamental wavelength ($\lambda$ = 1064 nm) generated by a Nd:YAG laser. During the second phase, to reduce the mean size of the as-obtained nanoparticles and properly adjust the size distribution, the water colloids were additionally irradiated by ultraviolet harmonics (355 nm and 266 nm) from the same laser source. The effect of the key laser parameters - wavelength, fluence and laser exposure time - upon the nanoparticles morphology was studied. The most suitable post-ablation treatment of initial colloids was obtained by consecutive irradiation with the third (λ = 355 nm) and the fourth (λ = 266 nm) harmonics of the fundamental laser wavelength. By using this approach synergistic effect between two mechanisms of light absorption by AgNPs was induced. As a result contaminant-free colloids of AgNPs with a size inferior to 10 nm and a quite narrow size distribution with a standard deviation of 1.6 nm were fabricated.

The toxic effect of the as-produced AgNPs on Gram-positive and Gram-negative bacteria and *Candida albicans* was explored. The most efficient action was reached against *Pseudomonas aeruginosa* and *Escherichia coli*.

Potential application of the synthesized AgNPs colloidal aqueous solutions with antimicrobial action as a non-invasive method for ocular infections prevention and treatment was proposed.

1. Introduction

The unique electronic [1, 2], optical [3, 4], magnetic [5-7] and catalytic properties [8, 9] of noble metals nanoparticles, essentially different from those of the bulk material, have attracted considerable research interest. It encompasses various fields, from fundamental sciences, such as physics (surface plasmon resonance (SPR) in visible and near UV regions or plasma discharge enhancement) [10], chemistry (catalysis) and biology (optical markers for diagnosis of various biological objects), to applied sciences, as design and development of a variety of electronic devices and sensors, and to industrial areas, as textile and cosmetics industry.

In what concerns silver particles of nanometer size (AgNPs), besides being nontoxic, they are known to suppress or destroy various types of bacteria [11]. Thus, they have provoked particular and wide-ranging interest for medicinal uses, e.g., improving the anti-bacterial effect of antibiotics [12], or as components in unguents for skin application [13]. Their size is smaller than that of human cells by several orders of magnitude, which is opening unique opportunities for biochemical interactions, either on the cells' surface or within cells [14]. Elechiguerra et al. [15] described AgNPs interacting with the HIV-1 virus thus suppressing in vitro its attachment to cells. Furthermore, silver nano-sized particles were found to inhibit the process of forming advanced glycation end-products, which could alleviate the adverse effects of diabetes [16]. Recently, these particles' beneficial capability in treating malignancies, and in diagnostic and probing were recognized [17].

This list of applications of silver nanoparticles, emphasizing their use in medicine, is relevant to the thematic orientation of the article. Their application in other areas is much wider, e.g.: a) detection by surface enhanced Raman spectroscopy (SERS) of ultra-low concentration of some molecules attached to AgNPs [18,19]; b) microelectronics; c) sunscreens and cosmetics [20,21]; d) clothing [22,23], etc. The increased interest to AgNPs due to their extremely wide range of applications led to development of various methods for their fabrication. In general, they can be classified as chemical [24], photochemical [25], electrochemical [26], sonochemical [27], biological [28], or physical [29]. The potential toxicity of the residual ions, as byproducts in colloids produced by various chemical methods, stays for the common disadvantage in respect with the prospect for further biological applications. The inability to fabricate large quantities of AgNPs is the major limitation of biological synthesis. Vacuum conditions and very high temperature associated with expensive equipment are connected as a rule with the use of physical methods. In this context, one promising physical method for fabrication of AgNPs colloids for medical applications is pulsed laser ablation under liquids (PLAL) [30]. A major advantage of this method is the fabrication of contamination-free colloids by using a simple and low-cost processing setup. The main disadvantage of PLAL approach compared to the other fabrication methods is the wider size distribution of produced nanoparticles.

The object of this study is fabrication of Ag nanocolloids for drug delivery in ophthalmology. As known, the passage of drugs in the form of eye drops to the posterior segment of the eye is severely restricted by eye barriers. Several ways of delivering drugs to the eye were currently being used: topical infusion and sub-conjunctival injection, trans-scleral drug delivery, intra-vitreal injection, sub-retinal injection [31,32]. The above-mentioned invasive methods may sometimes lead to post-administration complications, such as intraocular infections and retinal detachment. Nanoparticle delivery systems can reduce the incidence of injections and intraocular complications. In this case, the nanoparticles size is a key factor for ocular drug delivery efficiency. Bisht et al. [33] showed that the size of scleral water channels/pores is about 30 nm to 300 nm. Particles smaller than 20 nm can penetrate the corneal endothelium, which is the first eye barrier [34]. A narrow size distribution of NPs with sizes of less than 10 nm is an important requirement for NPs to serve as successful blood-retinal barrier transporters in eye drug delivery. This condition can be implemented via irradiating already-fabricated colloids with laser light under optimized parameters (wavelength, fluence, time of laser exposure) [35,36]. Such an approach has been employed mostly in laser synthesis of gold NPs to modify both their size-distribution and morphology [37-40]. The irradiation procedure could be conducted by either nano-, pico- or femtosecond lasers [41-44], taking into account the corresponding type of laser-mater interaction in nanoscale range.

We report on a simple approach for fragmentation of silver NPs by post-laser ablation treatment of water Ag nanocolloids via consecutive laser irradiation with 355 nm and 266 nm nanosecond pulses generated by the Nd:YAG laser source. Using this approach, for the first time, to the best of our knowledge, contaminant-free AgNPs with size below 10 nm and very narrow size distribution, namely, with mean diameter of 5.4 nm and a standard deviation of 1.6 nm, were fabricated. The AgNPs produced completely conform to the ophthalmology demands for blood-retinal barrier transporters in eye drug delivery. Our investigation by microbiological assay established strong antibacterial and antifungal activity of the newly fabricated Ag nanoparticle colloids. The potential application of the synthesized AgNPs colloidal aqueous solutions with antimicrobial action was proposed as a non-invasive method for ocular infections prevention and treatment.

## 2. Materials and Method

Our studies were organized in two steps: i) synthesis of AgNPs in water solution by multiple laser ablation of a silver target under water with the fundamental wavelength of a Nd:YAG source to produce the initial colloids; ii) induce controlled modifications of AgNPs by next irradiation with 355 and 266 nm harmonics from the same laser source.

The experiments were conducted as follows:

### 2.1. Ag nanoparticle colloids: fabrication in water – "first phase"

A pure silver target was submitted to multiple Nd:YAG laser irradiation at $\lambda = 1064$ nm with 15 ns laser pulses at a frequency repetition rate of 10 Hz. The target was placed on the bottom of a plastic beaker and immersed in double-distilled water (0.8 µS m$^{-1}$) at 6 mm below surface. 1800 subsequent laser pulses were applied for synthesis of AgNPs in solution. The total ablation time was three minutes at a fluence of 24.7 J cm$^{-2}$, Tab. 1. The laser beam was focused through a lens ($f = 22$ cm) perpendicularly to a disc-shaped silver target. A separation distance of 18 cm was chosen. The experimental setup employed to fabricate the initial AgNPs colloids is presented schematically in Fig. 1 and is similar to the one used in Nikov et al. [45]. In the present experiment the target was mounted on a stepper-motor computer-controlled *x-y* table. Thus, a precise control of the target translation motion was ensured and a virgin surface was exposed to each subsequent laser shot.

| Experimental phase | Mode | Laser wavelength [nm] | Laser fluence [J/cm²] | Time of exposure [min] |
|---|---|---|---|---|
| I |  | 1064 | 20.0 | 0.83 |
|  |  |  | 21.0 | 0.67 |
|  |  |  | 23.7 | 0.67 |
|  |  |  | 24.7 | 0.67 |
|  |  |  | 26.8 | 0.16 |
| II | 1 | 355 | 0.059 | 85 |
|  |  |  | 0.068 | 5 |
|  | 2 | 266 | 0.023 | 55 |
|  | 3 | 355 | 0.059 | 85 |
|  |  |  | 0.068 | 5 |
|  |  | 266 | 0.023 | 20 |

**Tab. 1:** Numerical values of important technological parameters of the laser beam used in the different experimental phases.

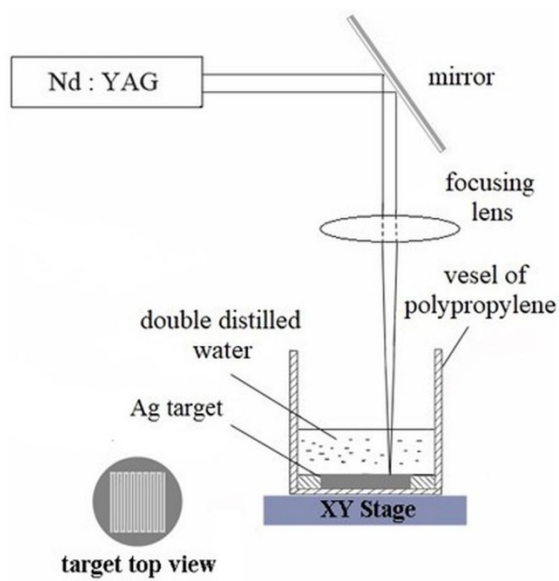

**Fig. 1**: Experimental set-up employed to fabricate the initial water colloids of AgNP

## 2.2. Fragmentation of initial colloids by third and fourth harmonics of Nd:YAG laser source – "second phase"

In the second phase of the experimental procedure, in order to fragment the silver nanoparticles the initial colloids were irradiated with the third ($\lambda$ = 355 nm) and the fourth ($\lambda$ = 266 nm) harmonics generated by the same Nd:YAG laser source.

These two wavelengths were selected since they are located close to the plasmon and to the inter-band absorption [46] peaks of AgNPs, respectively. The experimental set-up was similar that one described in Nikolov et al. [35]. The colloids were stored in a container rotating at a frequency of 4.3 rpm. The unfocused laser beam was directed normally to the top surface of colloids. The experiments of the second phase were conducted by irradiating the initial colloid under three different modes. In the *first two modes* the colloid was irradiated with either the wavelength of 355 nm, or with the wavelength of 266 nm, only. According to our studies, both modes reduce the AgNPs mean size and narrow their size distribution in a different way, but not appropriate for the medical application discussed above.

Combined irradiation was therefore applied in the *third mode*, by using sequentially the wavelengths of 355 nm and 266 nm, respectively. To the best of our knowledge, this procedure was for the first time applied to reach the main goal of the treatment – fabricating the colloid suitable for use in ophthalmology (Patent application) [47]. The experimental parameters (laser wavelength and fluence, and time of laser exposure) used in the second phase were collected in Tab. 1 and were proved to be optimal over the entire range of experimental conditions. The total irradiation time was determined based upon the effect of laser treatment on AgNPs optical transmission spectra obtained during the previous step. The optical transmission of colloids was measured after each step of irradiation lasting 5 min according to the procedure described in Nikolov et al. [35]. It showed the evolution of the transmission spectra as the time of laser exposure was raised.

### 2.3. Physical-chemical characterization of Ag nanoparticles

The optical transmission spectra of AgNPs water colloids in the UV/Vis range (300-900 nm) were recorded by an Ocean Optics HR 4000 spectrometer. The size and the morphology of AgNPs were visualized by transmission electron microscopy (TEM). The images were acquired with a JEOL JEM 2100 at an accelerating voltage of 200 kV. High-resolution transmission electron microscopy (HRTEM) and selected area electron diffraction (SAED) were used to characterize the nanoparticles microstructure and their phase composition.

### 2.4. Ag nanoparticles: Microbiological assay

The antibacterial and antifungal activity of the newly fabricated Ag nanoparticle colloids were assessed by using Gram-positive and Gram-negative bacteria, and Candida albicans. The following strains were used: *Staphylococcus aureus* strain 29213 and *Escherichia coli* strain 35218 from an

American Collection of Cell Cultures (ATCC); *Pseudomonas aeruginosa* strain 1390 and *Candida albicans* strain 74 from the collection of the Stefan Angeloff Institute of Microbiology, Bulgarian Academy of Sciences. Suspensions of the respective microorganisms with a concentration of $1 \times 10^6$ CFU mL$^{-1}$ were prepared. The experiments were conducted with an incubation mixture containing equal volumes of suspensions of microorganisms and AgNPs. The mixture was poured in 12-well polystyrene plates, continuously shaken, and sampled at pre-defined intervals (0, 2, 5 and 24 hours) in order to count the number of viable microorganisms. Ten-fold diluted solutions of the incubation mixtures in a nutrient medium (Tryptic soy agar, Oxoid) were seeded and the colonies-forming units (CFUs) were counted after culturing for 24 hours at 37 ºC.

To confirm the results, the experiments were repeated three times, with numerical values presented As an average value +/- mean standard deviation (SD). The difference between two means was determined by a two-tailed unpaired Student's test.

## 3. Results and Discussion

### 3.1. Photo fabrication of Ag nanocolloids

In Fig. 2 were collected the typical TEM micrographs (a), distribution histograms (b) and corresponding SAED patterns (c) of the AgNPs obtained after pulsed laser ablation and following irradiation, respectively, in water environment. They were organized in the respective order: Ag colloids after the *first phase* (I) and after *second phase* (II-IV), which corresponded to the three modes of UV irradiation introduced above.

Fig. 2 I-a, is indicative for spherical and spherical-like AgNPs shape of initial colloid prepared during the ablation process in the first experimental phase. Also, formation of aggregates was observed. Ag grains with a mean size of ~ 20 nm (Fig. 2 I-b) exceeds the most probably by a factor of more than two the upper limit of the proper size needed for the specific medical application [33-34].

As expected, the application of the second phase resulted in fragmentation of AgNPs to a mean size of 15.2 nm, 8.43 nm after irradiation with 355 nm, 266 nm only and 5.4 nm after the consecutive irradiation with wavelengths of 355 nm and 266 nm for the three irradiation modes, respectively, Fig. 2II-IV. SD decreased significantly in all cases after the second phase (3.5, 6.44 and 1.6 nm) in respect with the first phase (15.4 nm). It noteworthy, that the smallest mean size and SD values were observed after the consecutive irradiation with wavelengths of 355 nm and 266 nm (the third irradiation mode), Fig. 2-IV.

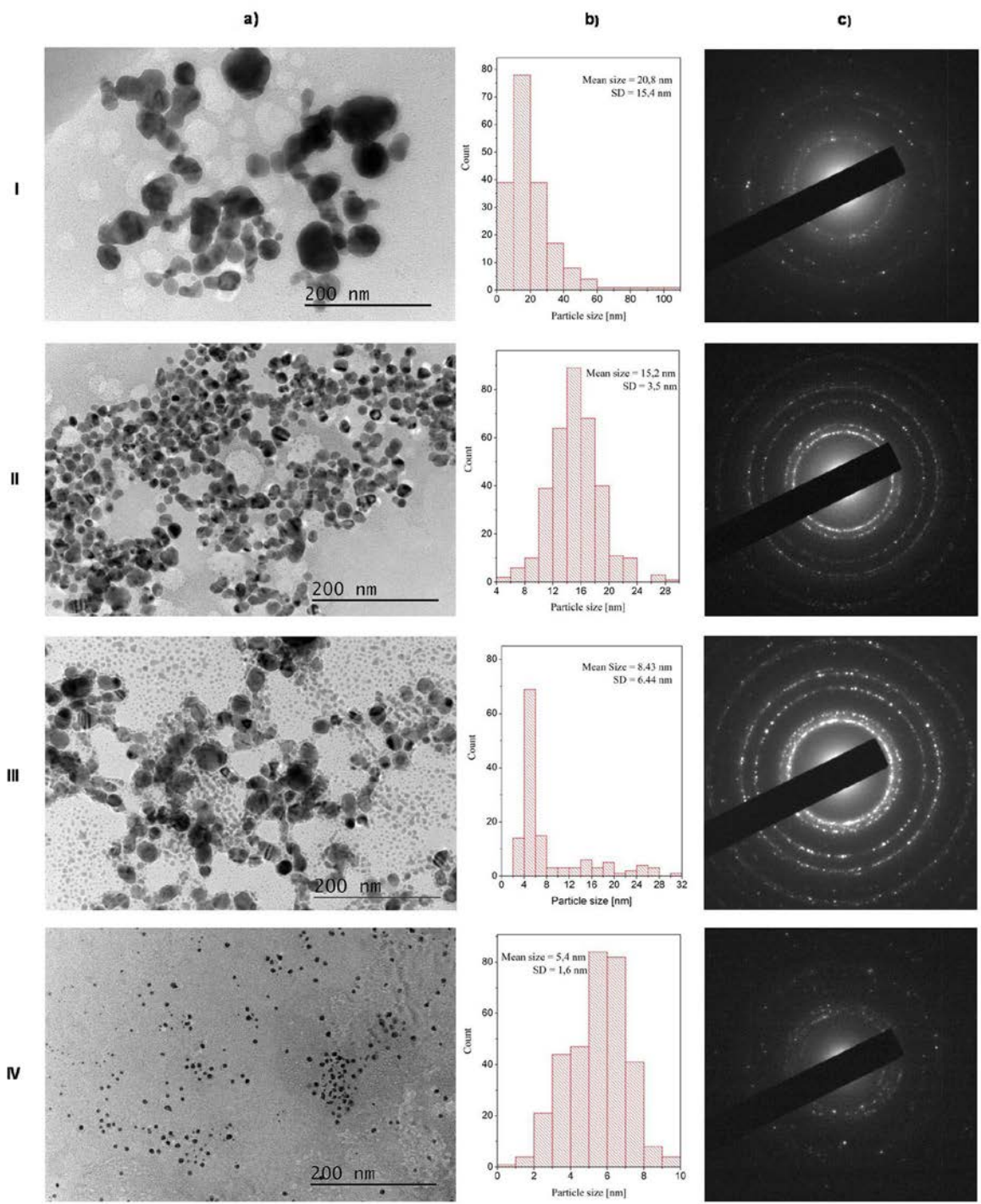

**Fig.2**: a) Typical TEM micrographs, b) Particles' size distribution histograms, c) corresponding SAED patterns of Ag colloids produced after the first phase (I) – initial colloid, and the second phase (II) with 355 nm– after 85 min, 266 nm (III) – after 40 min, and consecutive application of 355 (after 85 min) and 266 nm (after 15 min) wavelengths laser pulses, respectively - (IV).

An important note is that all SAED patterns confirmed the unique presence of Ag in a polycrystalline state, Fig. 2c.

The optical transmission of the colloidal solutions produced under different irradiation conditions was investigated. The results are displayed in Fig. 3-5.

One can see in Fig. 3 the transmission behavior through colloidal solutions after completion the first phase (a) and second phase with $\lambda = 355$ nm and fluence of 0.059 J cm$^{-2}$ for 30 (b), 50 (c), 70 (d) and 85 (e) minutes of irradiation time. For comparison, the fluence was raised up to 0.068 J cm$^{-2}$ and the colloid was laser exposed 5 min more (complete irradiation of 90 min) in Fig. 3f. From the examination of Fig. 3, the following observations were in order:

i) The transmission was the lowest in case of the initial colloid produced in the first phase.

ii) It gradually increased with the number of subsequent laser pulses (total time duration) applied under the second phase.

iii) Saturation appeared when duration exceeded 85 min, even when laser fluence was increased to 0.068 J cm$^{-2}$ (Fig. 3f).

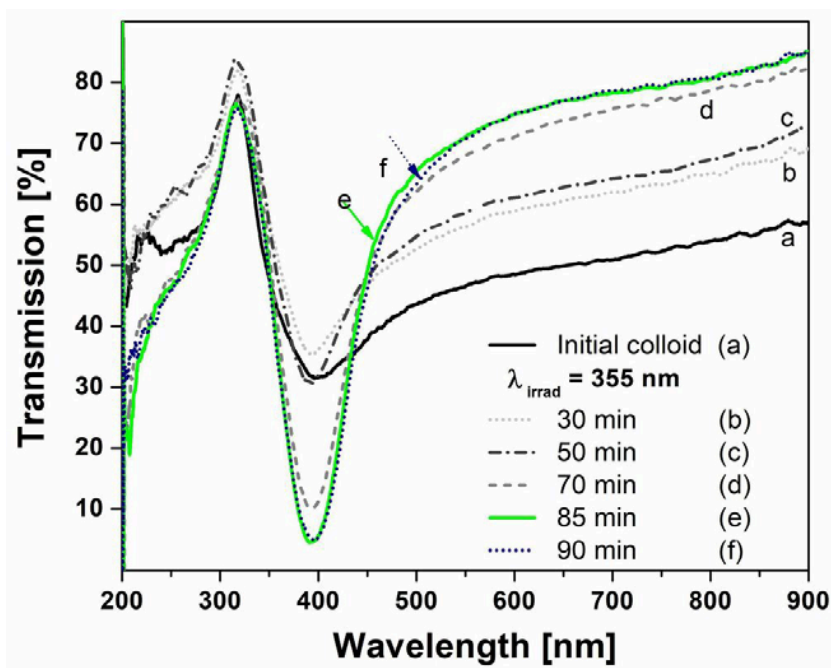

**Fig. 3**: Optical transmission spectra of AgNPs aqueous colloids: a) initial colloid after laser ablation with 1064 nm and next irradiation with 355 nm after: b) 30 min, c) 50 min, d) 70 min, e) 85 min, f) 90 min, respectively.

A prominent band within the range of 300 - 600 nm with a minimum at ~ 390 nm is present in all spectra. This band should the most probably assigned to surface plasmon resonance (SPR) associated to the free electrons absorption in silver nanoparticles [48]. It confirms the correlation between SPR characteristics and the mean size and distribution of NPs [49]. The increase/decrease of nanoparticles mean size can be accounted in terms of a red shift/blue shift of SPR band. After completion of each irradiation step, the SPR band progressively narrowed, the value of its minimum decreased and the spectrum shoulder extension towards high-wavelengths region rose. This behavior was caused by size-redistribution in nanoparticles ensemble, which might be due to relatively low values of laser fluence used in our experiments, in accordance with "particle heating-melting-evaporation" model, proposed by Takami et al. [50]. It is well-known, the NPs absorption depends on their geometric cross-section, i.e. size conditioned by dumping and retardation effects and the SPR position in respect with incident irradiation wavelength. Moreover, the smaller NPs in water colloids are characterized by a higher heat-dissipation rate to the surrounding water [43,49,51] because of increased surface/volume ratio.

Laser treatment of colloids initiate two concurrent processes: fragmentation of big nanoparticles, and concomitant with aggregation and fusion of small ones. The energy of the electrons excited by SPR is transmitted by electron-phonon coupling to lattice. When temperature reached the melting point, latent heath should be further delivered to get NPs complete melting. The temperature excursion continued to evaporation and boiling point, which also needed latent heat compensation and resulted in reduction of AgNPs mean size, by selective fragmentation of the biggest ones. On the other hand, fusion of NPs could proceed by aggregation and melting.

The equilibrium between these two processes, fragmentation and fusion, depends on the laser pulses energy and duration of irradiation and is reflected in NPs mean size and size distribution. The reduction trend of these characteristics of the ensemble of the NPs continued therefore up to 85-min irradiation duration, as evidenced by the optical transmission spectra in Fig. 3e. During the pulsed laser exposure, the fragmentation tendency obviously dominated the particle-size redistribution evolution. However, the next increase of the irradiation time and/or the laser fluence led to a widening of the plasmon band in the transmission spectrum. This could be the result of enhanced aggregation and fusion of AgNPs.

The application of the second mode of illumination was therefore extended in order to preserve the photo-fragmentation predominance. It was conducted in analogous way of the previous one ($\lambda$ = 355 nm), except the use of more energetic photons coming with the fourth harmonic ($\lambda$ = 266 nm - 4.66 eV, as against 3.49 eV for 355 nm). Very important, SPR absorption was replaced in this case by an absorption process due to an inter-band transition [46].

The results obtained by optical transmission spectroscopy after the two steps of laser irradiation process, i.e. initial colloids generation by 1064 nm, followed by supplementary fragmentation of AgNPs under 266 nm irradiation for different irradiation times are displayed in Fig. 4. One can notice in Fig. 4 a similar tendency to that observed in Fig. 3, i.e. a progressive narrowing of the SPR band and raise of the spectrum wing to the long-wavelengths region. Nevertheless, saturation appeared again after about 50 min of irradiation only this time. It is followed by a reversal trend of NPs size increase and distribution widening.

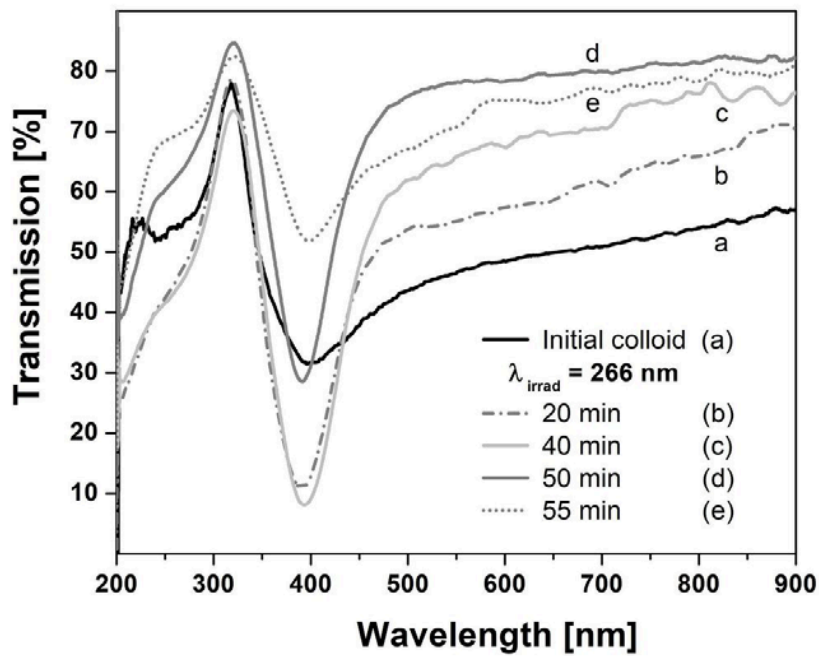

**Fig. 4:** Optical transmission spectra of AgNPs aqueous colloids: a) initial colloid after laser ablation with 1064 nm and after next laser irradiation with 266 nm for: b) 20 min, c) 40 min, d) 50 min, e) 55 min, respectively.

The best AgNPs colloids were obtained after 40 min of irradiation (Fig. 4, spectrum "c") but their mean size were still inappropriate for ophthalmologic applications.

Another feature that should be mentioned when comparing TEM images obtained in both measurement modes was the appearance of two populations of nanoparticles – bigger and smaller. This process was barely noticeable in the first irradiation mode (Fig. 2-II), but was very pronounced in the second one (Fig. 2-III), which reflected in the relatively wide nanoparticle size distribution in this case. The standard deviation value (6.44 nm) of the AgNPs mean size in the colloid obtained after irradiation with only 266 nm (second mode) is nearly twice the standard deviation (3.5 nm) of the average size of nanoparticles obtained after irradiation with only 355 nm (first mode).

One may conclude, that the first two modes of laser irradiation reduced AgNPs mean size and narrowed their size distribution in a different way, without being able however to fully meet the specific requirements for medical use in ophthalmologic therapy.

That is why we used a combined treatment by the third ($\lambda = 355$ nm) and the fourth ($\lambda = 266$ nm) harmonics, successively applied for irradiation of the initial colloid in the third mode of the second phase of the experimental procedure. It was organized in analogy with the first two modes. The initial colloid was irradiated during the first 85 minutes with $\lambda = 355$ nm at the same laser energy as in Fig. 3. The change of the optical transmission spectrum of the initial colloid after 85 and 90 minutes of irradiation with 355 nm is seen in Fig. 5.

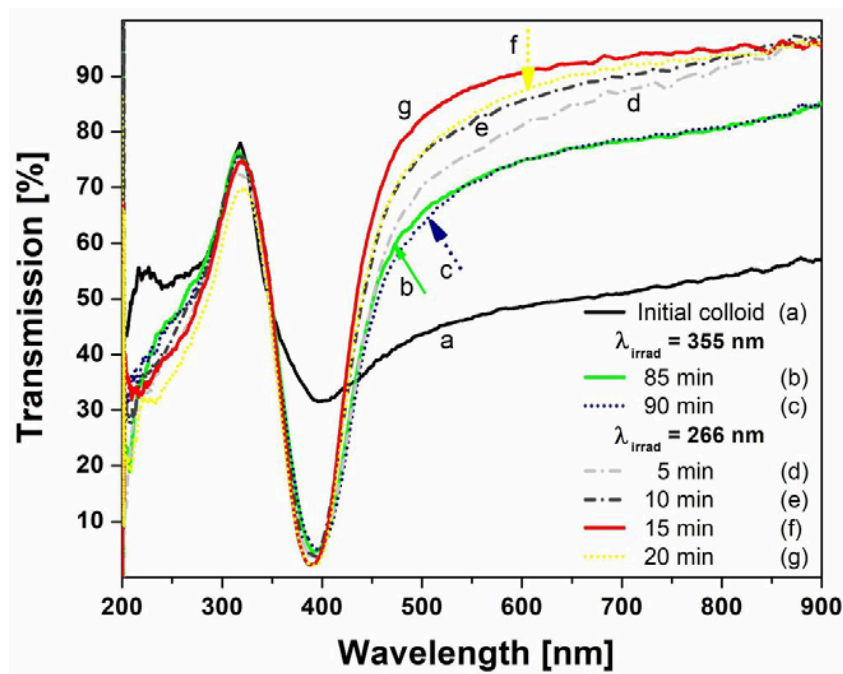

**Fig. 5**: Optical transmission spectra of AgNPs aqueous colloids: a) initial colloid after laser ablation with 1064 nm and the consecutive laser irradiation with 355 nm after: b) 85 min, c) 90 min, followed by 266 nm for: d) 5 min, e) 10 min, f) 15 min, g) 20 min, respectively.

The irradiation was continued with 266 nm for different durations. A certain narrowing was observed after 5 min and 10 min irradiation time of the plasmonic band, together with a slight blue-shift of its minimum. The latter is a sign of a reduction of the mean particle size as an effect of fragmentation of large particles. Reducing the number of large particles should lead to an increase in the transmission values in the spectrum long wavelength region, which was actually observed. The maximum effect was reached after 15 min laser irradiation (Fig. 5, spectrum "f").

The further irradiation prolongation caused the reversal trend, as evidenced by the optical transmission spectrum after 20 min (Fig. 5g). The average size of the silver nanoparticles produced in the third mode and its corresponding standard deviation (5.4 nm and 1.6 nm, respectively) were the smallest ones compared with these ones of the AgNPs produced in the first and second modes. This proved that a synergistic effect was reached with the third mode of irradiation between the sequential actions of the two above mentioned mechanisms of absorption, in respect with the average size and size distribution of AgNPs.

Very important, TEM examination showed that generated nanoparticles preserved the spherical shape in this case too (Fig. 2a).

In conclusion, the *third mode* completely meets the aim of the treatment, i.e. in preparation of silver nanocolloids suitable for use in ophthalmology. Thus, the fabrication of nearly monodisperse Ag nanoparticles with an average size of 5.4 nm allow the drugs in the form of eye drops to pass through most of eye barriers, such as scleral water channels/pores, and can penetrate the corneal endothelium [33-34].

### 3.2. Antimicrobial assay

The results reported in this Section were obtained with AgNPs of ~ 5.4 nm mean size, produced with the third irradiation mode, under optimum conditions.

Our studies showed that the AgNPs were the most effective against *Pseudomonas aeruginosa*. They reduced the number of the bacteria by four logs (10 000 times, i.e lower than the minimum level of viable microorganisms, (Fig. 6) after two hours.

Then the number of viable bacteria dropped fast and after five hours they were completely inactivated. A very effective action was obtained against *Escherichia coli* for which full inactivation was also observed after five hours. The process of inactivation ran rapidly between the second and the fifth hour (Fig. 7).

Ag nanocolloids reduced the number of *Staphylococcus aureus* by two logs (100 times) after five hours and completely inactivated the bacteria after 24 hours (Fig. 8).

The antifungal action against *Candida albicans* actively started after five hours. The number of fungi was reduced by 1.5 logs. The process continued for 24 hours, when fungi were completely inactivated (fig. 9).

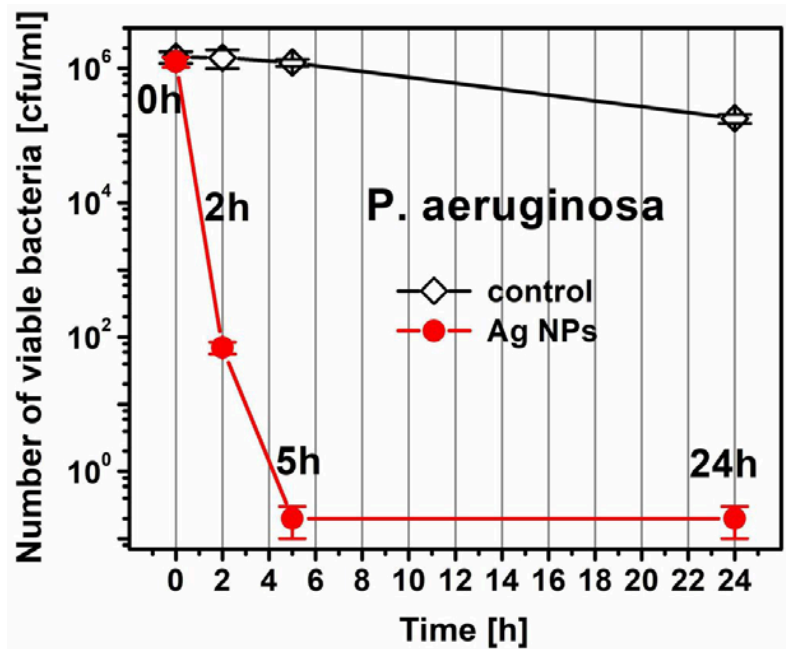

**Fig. 6**: Antibacterial action of Ag NPs produced by the third mode of irradiation against *Pseudomonas aeruginosa*. The viable bacteria number decreased 10000 times after the 2-nd hour, while full inactivation was reached after the 5-th hour.

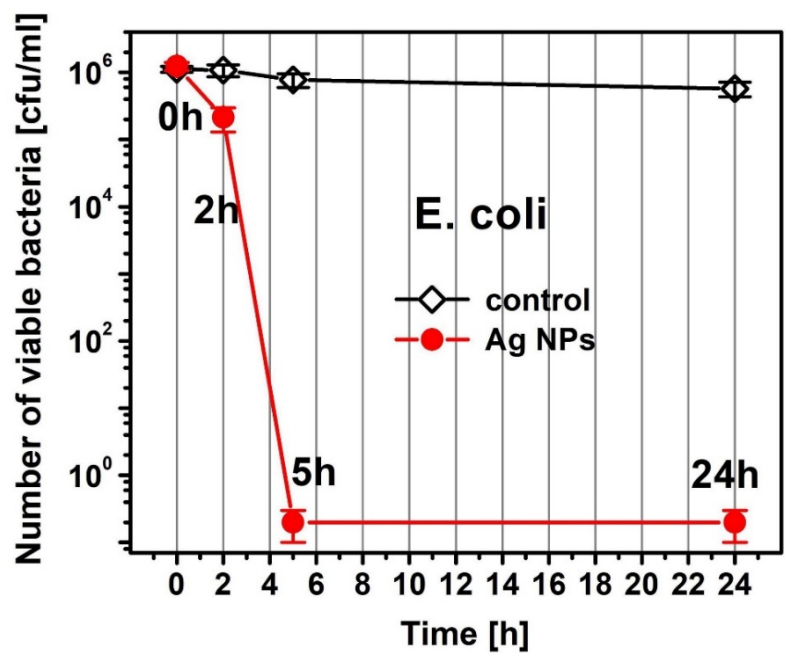

**Fig. 7**: Antibacterial action of Ag NPs produced by the third mode of irradiation against *Escherichia coli*. Full inactivation was obtained after the 5-th hour.

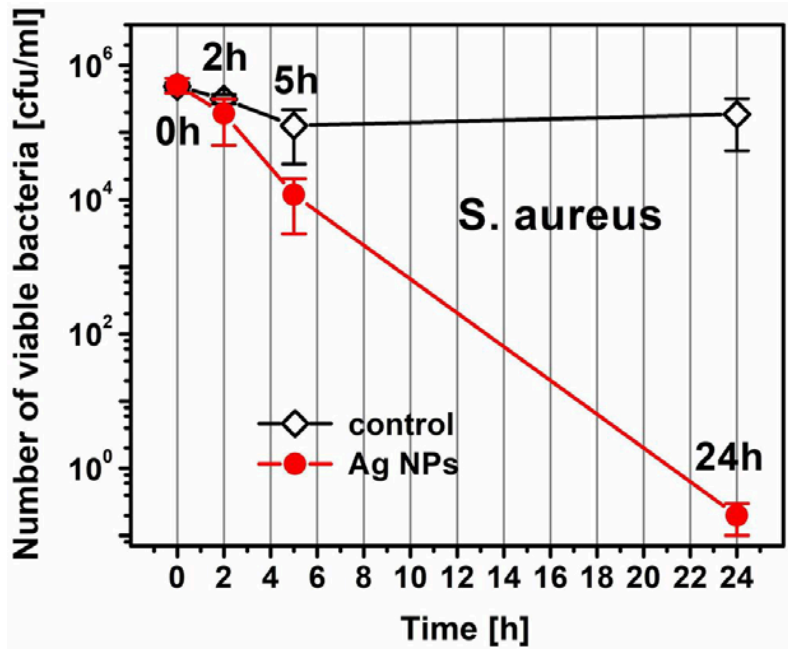

**Fig.8**: Antibacterial action of Ag NPs produced by the third mode of irradiation against *Staphylococcus aureus*. The inactivation process started after 5 h. The number of viable bacteria decreased 100 times (viable microbes are below the minimum). Full inactivation was reached after 24 hours.

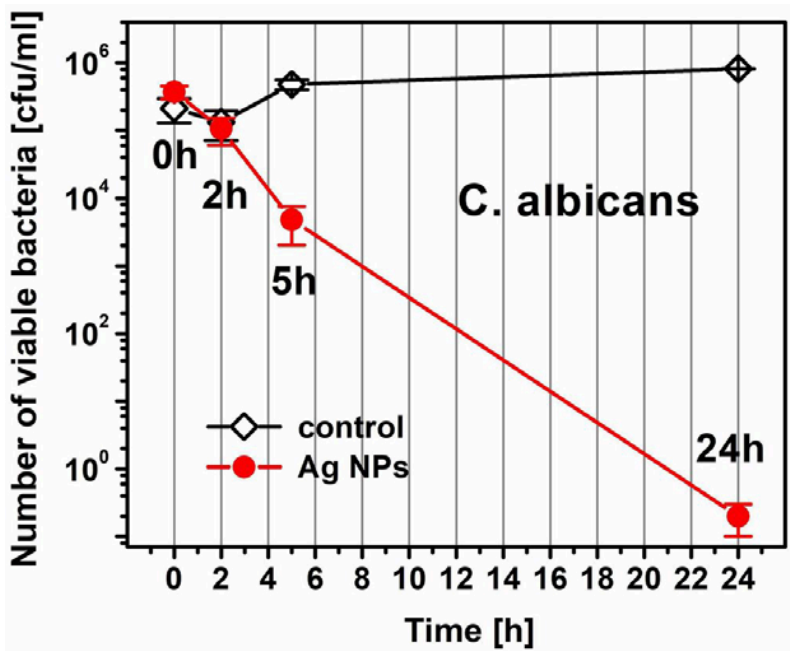

**Fig. 9:** Antibacterial action of Ag NPs produced by the third mode of irradiation against *Candida albicans*. Full inactivation was obtained after 24 hours. The number of viable fungi decreased 31 times after 5 hours.

Antibacterial power of silver has been widely documented [52-54], although the exact mechanism causing it has not been completely understood yet. It is believed that it might derive from the inactivation of enzymes essential for the respiratory chain of the pathogen, or by generating hydroxyl radicals [55]. They cause, in turn, pathogens annihilation.

## 4. Conclusions

Pulsed laser ablation under different irradiation regimes was applied to photofabricate Ag nanocolloids with controlled mean size and distribution, suitable for potential ophthalmological applications.

To the best of our knowledge, a consecutive application of the third and the fourth harmonics after the fundamental wavelength of a Nd:YAG laser source was for the first time investigated to this purpose. It proved efficient to initiate synergistic effects between consecutive application of the fundamental and the harmonics (third and fourth ones) to the same Ag nanocolloidal population. Chemical-free AgNPs with a mean size of 5.4 nm and a very narrow size distribution (standard deviation of ~ 1.6 nm) was fabricated under the optimal experimental conditions - consecutive laser irradiation of the water colloids of silver nanoparticles with 355 nm and 266 nm.

Antibacterial action of AgNPs was explored with good results against *Pseudomonas aeruginosa*, *Escherichia coli*, *Staphylococcus aureus* and *Candida albicans*.

Ag nanocolloids proved most efficient against *Pseudomonas aeruginosa* and *Escherichia coli* bacteria, for which full apoptosis was reached after five hours.

The small size of the AgNPs allows for an expected efficient cure of intraocular infections.

One should emphasize upon the key role of laser pulse energy for narrowing AgNPs size distribution as a result of laser multiple irradiation.

Our results support the efficient green photofabrication of ultrafine monodisperse Ag nanoclusters by a multistep laser procedure for applying them as non-invasive method of prevention and cure of intraocular infections of any sort.

One should mention that all irradiations - i.e. for both initial generation of Ag nanocolloids but also for the next adjustment of their size and size distribution (1064 vs. 355 and 266 nm) - were generated by the same Nd:YAG laser source. This means a procedure easy to informatize and automize for low cost exploitation.

**Acknowledgements**

Research equipment of distributed research infrastructure INFRAMAT (part of Bulgarian National roadmap for research infrastructures) supported by Bulgarian Ministry of Education and Science under contract D01-284/17.12.2019 was used in this investigation. This work was supported by the National Science Fund, Ministry of Education and Science, Sofia, Bulgaria (Contract № KP-06-H 38/13, 06.12.2019, "Development of Biophotonics methods as the basis of Oncology Theranostics - 2").

Romanian co-authors acknowledge with thanks the Core Programme 16 N/2019. Bilateral cooperation under project "Laser-assisted processing of materials and their characterization for high technology applications" (2020-2022) between Romanian and Bulgarian Academies of Sciences are also acknowledged.

**References**

[1] D. Chen, X. Qiao, X. Qiu, J. Chen, "Synthesis and electrical properties of uniform silver nanoparticles for electronic applications", Journal of Materials Science, 44 (2009) 1076-1081.

[2] A.H. Alshehri, M. Jakubowska, A. Młożniak, M. Horaczek, D. Rudka, C. Free, J.D. Carey, Enhanced electrical conductivity of silver nanoparticles for high frequency electronic applications, ACS Appl. Mater. Interfaces 4, 12, 7007-7010. doi:10.1021/am3022569.

[3] S. Eustis, M.A. El-Sayed, Why gold nanoparticles are more precious than pretty gold: Noble metal surface plasmon resonance and its enhancement of the radiative and nonradiative properties of nanocrystals of different shapes, Chemical Society Reviews 35 (2006) 209–217. https://doi.org/10.1039/B514191E.

[4] C. Noguez, Surface plasmons on metal nanoparticles: the influence of shape and physical environment, The Journal of Physical Chemistry C, 111 (2007) 3806-3819. doi:10.1021/jp066539m.

[5] J.S. Garitaonandia, M. Insausti, E. Goikolea, M. Suzuki, J.D. Cashion, N. Kawamura, H. Ohsawa, I. Gil de Muro, K. Suzuki, F. Plazaola, Chemically induced permanent magnetism in Au, Ag, and Cu nanoparticles: localization of the magnetism by element selective techniques, Nano Lett., 8 (2008) 661-667. doi: 10.1021/nl073129g.

[6] H. Le Trong, K. Kiryukhina, M. Gougeon, V. Baco-Carles, F. Courtade, S. Dareys, P. Tailhades, Paramagnetic behaviour of silver nanoparticles generated by decomposition of silver oxalate, Solid State Sciences, 69 (2017) 44-49. doi: 10.1016/j.solidstatesciences.2017.05.009.

[7] Y. Yamamoto, T. Miura, Y. Nakae, T. Teranishi, M. Miyake, H. Hori, Magnetic properties of the noble metal nanoparticles protected by polymer, Physica B: Condensed Matter 329–333, Part 2 (2003) 1183-1184. doi:10.1016/s0921-4526(02)02102-6.

[8] K. de O. Santos, W.C. Elias, A.M. Signori, F.C. Giacomelli, H. Yang, and J.B. Domingos, Synthesis and Catalytic Properties of Silver Nanoparticle–Linear Polyethylene Imine Colloidal Systems, J. Phys. Chem. C 116 (7) (2012) 4594–4604. doi: 10.1021/jp2087169.

[9] Y.S. Jeong, J.-B. Park, H.-G. Jung, J. Kim, X. Luo, J. Lu, L. Curtiss, K. Amine, Y.-K. Sun, B. Scrosati, and Y.J. Lee, Study on the Catalytic Activity of Noble Metal Nanoparticles on Reduced Graphene Oxide for Oxygen Evolution Reactions in Lithium–Air Batteries, Nano Lett., 15 (7) (2015) 4261–4268, doi: 10.1021/nl504425h.

[10] B.Zhao, I.Aravind, S.Yang, Z. Cai,Y. Wang, R. Li, S. Subramanian, P. Ford, D. R. Singleton, M. A. Gundersen, S. B. Cronin, Nanoparticle-Enhanced Plasma Discharge Using Nanosecond High-Voltage Pulses, The Journal of Physical Chemistry C (2020), DOI: 10.1021/acs.jpcc.9b12054

[11] K.H. Cho, J.E. Park, T. Osaka, S.G. Park, The study of antimicrobial activity and preservative effects of nanosilver ingredient, Electrochimica Acta 51 (2005) 956-960, doi:10.1016/j.electacta.2005.04.071.

[12] A.R. Shahverdi, A. Fakhimi, H.R. Shahverdi, S. Minaian, Synthesis and effect of silver nanoparticles on the antibacterial activity of different antibiotics against Staphylococcus aureus and Escherichia coli, Nanomedicine: Nanotechnology,Biology, and Medicine 3 (2007) 168-171. doi:10.1016/j.nano.2007.02.001.

[13] S. Kokura, O. Handa, T. Takagi, T. Ishikawa, Y. Naito, T. Yoshikawa, Silver nanoparticles as a safe preservative for use in cosmetics, Nanomedicine: Nanotechnology, Biology and Medicine, 6(4) (2010) 570–574. doi:10.1016/j.nano.2009.12.002.

[14] J. Natsuki, T. Natsuki, Y. Hashimoto, A Review of Silver Nanoparticles: Synthesis Methods, Properties and Applications, Int. J. Mater. Sci. Appl. 4(5) (2015) 325–332. doi: 10.11648/j.ijmsa.20150405.17.

[15] J.L. Elechiguerra, J. Burt, J.R. Morones, A. Camacho-Bragado, X. Gao, H.H. Lara,

M.J. Yacaman, Interaction of silver nanoparticles with HIV-1, J Nanobiotechnol 3, 6 (2005). https://doi.org/10.1186/1477-3155-3-6.

[16] J.M. Ashraf, M.A. Ansari, H.M. Khan, M.A. Alzohairy, I. Choi, Green synthesis of silver nanoparticles and characterization of their inhibitory effects on AGEs formation using biophysical techniques, Scientific Reports 6(1) 20414 (2016), doi:10.1038/srep20414.

[17] X.-F. Zhang, Z.-G. Liu, W. Shen, S. Gurunathan, Silver Nanoparticles: Synthesis, Characterization, Properties, Applications, and Therapeutic Approaches, Int. J. Mol. Sci. 17(9) (2016) E1534. doi: 10.3390/ijms17091534.

[18] S.R. Emory, S. Nie, Near-Field Surface-Enhanced Raman Spectroscopy on Single Silver Nanoparticles, Anal. Chem. 69, 14 (1997) 2631-2635. doi: doi:10.1021/ac9701647.

[19] S.R. Emory, W.E. Haskins, S. Nie, Direct Observation of Size-Dependent Optical Enhancement in Single Metal Nanoparticles, J. Am. Chem. Soc. 120, 31 (1998) 8009-8010. doi: 10.1021/ja9815677.

[20] N. Tyagi, S.K. Srivastava, S. Arora, Y. Omar, Z. M. Ijaz, A. AL-Ghadhban, S.K. Deshmukh, J.E. Carter, A.P. Singh and S. Singh, Comparative analysis of the relative potential of silver, zinc-oxide and titanium-dioxide nanoparticles against UVB-induced DNA damage for the prevention of skin carcinogenesis, Cancer Lett. 383(1) (2016) 53–61. doi:10.1016/j.canlet.2016.09.026.

[21] S. Gajbhiye, S. Sakharwade, Silver Nanoparticles in Cosmetics, Journal of Cosmetics, Dermatological Sciences and Applications, 6(01) (2016) 48-53. doi: 10.4236/jcdsa.2016.61007.

[22] F. Zhang, X. Wu, Y. Chen, H. Lin, Application of silver nanoparticles to cotton fabric as an antibacterial textile finish, Fibers and Polymers 10(4) (2009) 496-501. doi:10.1007/s12221-009-0496-8.

[23] H. Wigger, Case study: Clothing textiles with incorporated silver nanoparticles, Environmental Release of and Exposure to Iron Oxide and Silver Nanoparticles, Springer Vieweg, Wiesbaden, 2017. doi: 10.1007/978-3-658-16791-2.

[24] K. Gudikandula & S.C. Maringanti, Synthesis of silver nanoparticles by chemical and biological methods and their antimicrobial properties, Journal of Experimental Nanoscience 11(9) (2016) 714-721. doi:10.1080/17458080.2016.1139196.

[25] M. Zaarour, M. El Roz, B. Dong, R. Retoux, R. Aad, J. Cardin, C. Dufour, F. Gourbilleau, J.-P. Gilson, and S. Mintova, Photochemical Preparation of Silver Nanoparticles Supported on Zeolite Crystals, Langmuir, 30 (21) (2014) 6250–6256. doi:10.1021/la5006743.

[26] X. Zhang, E.M. Hicks, J. Zhao, G.C. Schatz, and R.P. Van Duyne, Electrochemical Tuning of Silver Nanoparticles Fabricated by Nanosphere Lithography, Nano Letters 5(7) (2005) 1503-1507. doi: 10.1021/nl050873x.

[27] J. Zhu, S. Liu, O. Palchik, Y. Koltypin, and A. Gedanken, Shape-Controlled Synthesis of Silver Nanoparticles by Pulse Sonoelectrochemical Methods, Langmuir 16(16) (2000) 6396-6399. doi: 10.1021/la991507u.

[28] ] S. Iravani, H. Korbekandi, S.V. Mirmohammadi, and B. Zolfaghari, Synthesis of silver nanoparticles: chemical, physical and biological methods, Res Pharm Sci 9(6) (2014) 385–406.

[29] J. Siegel, O. Kvítek, P. Ulbrich, Z. Kolská, P. Slepička, V. Švorčík, Progressive approach for metal nanoparticle synthesis, Mater. Lett. 89 (2012) 47–50. doi:10.1016/j.matlet.2012.08.048.

[30] G.W. Yang, Laser ablation in liquids: Applications in the synthesis of nanocrystals, Progress in Material Science, 52 (2007) 648-698. doi: 10.1016/j.pmatsci.2006.10.016.

[31] C.-H. Tsai, P.-Y. Wang, I-C. Lin, H. Huang, G.-S. Liu, and C.-L. Tseng, Ocular Drug Delivery: Role of Degradable Polymeric Nanocarriers for Ophthalmic Application, Int J Mol Sci 19(9) (2018) 2830. doi:10.3390/ijms19092830.

[32] V.K. Yellepeddi, S. Palakurthi, Recent advances in topical ocular drug delivery, J Ocul Pharmacol Ther 32(2) (2016) 67–82, doi:10.1089/jop.2015.0047.

[33] R. Bisht, A. Mandal, J.K. Jaiswal, I.D. Rupenthal, Nanocarrier mediated retinal drug delivery: Overcoming ocular barriers to treat posterior eye diseases, Wiley Interdisciplinary Reviews:Nanomed. Nanobiotechnol. 10(2) (2018) e1473, doi: 10.1002/wnan.1473.

[34] A. Rajapakshal, M. Fink, B.A. Todd, Size-dependent diffusion of dextrans in excised porcine corneal stroma, Mol Cell Biomech 12(3) (2015) 215–230. doi:10.3970/mcb.2015.012.215.

[35] A.S. Nikolov, R.G. Nikov, I.G. Dimitrov, N.N. Nedyalkov, P.A. Atanasov, M.T. Alexandrov, D.B. Karashanova, Modification of the silver nanoparticles size-distribution by means oflaser light irradiation of their water suspensions, Applied Surface Science 280 (2013) 55– 59. doi.org/10.1016/j.apsusc.2013.04.079.

[36] H. M. Momen, Effects of particle size and laser wavelength on heating of silver nanoparticles under laser irradiation in liquid, Pramana – J. Phys. 87: 26 (2016), doi:10.1007/s12043-016-1233-7.

[37] S. Inasawa, M. Sugiyama, and Y. Yamaguchi, Laser-Induced Shape Transformation of Gold Nanoparticles below the Melting Point:  The Effect of Surface Melting, J. Phys. Chem. B 109(8) (2005) 3104-3111. doi:10.1021/jp045167j.

[38] M. Honda, Y. Saito, N.I. Smith, K. Fujita, and S. Kawata, Nanoscale heating of laser irradiated single gold nanoparticles in liquid, Optics Express 19 (13) (2011) 12375-12383. doi: 10.1364/oe.19.012375.

[39] F. Mafuné, J-Y. Kohno, Y. Takeda, and T. Kondow, Dissociation and Aggregation of Gold Nanoparticles under Laser Irradiation, J. Phys. Chem. B, 105(38) (2001) 9050–9056, doi:10.1021/jp0111620.

[40] S. Link, M. El-Sayed, Spectral Properties and Relaxation Dynamics of Surface Plasmon Electronic Oscillations in Gold and Silver Nanodots and Nanorods, J. Phys. Chem. B 103(40) (1999) 8410–8426, doi:10.1021/jp9917648.

[41] S. Link & M.A. El-Sayed, Shape and size dependence of radiative, non-radiative and photothermal properties of gold nanocrystals, International Reviews in Physical Chemistry, 19:3 (2000) 409-453. doi: 10.1080/01442350050034180.

[42] D. Catone, A. Ciavardini, L. Di Mario, A. Paladini, F. Toschi, A. Cartoni, I. Fratoddi, I. Venditti, A. Alabastri, R.P. Zaccaria, and P. O'Keeffe, Plasmon Controlled Shaping of Metal Nanoparticle Aggregates by Femtosecond Laser-Induced Melting, J. Phys. Chem. Lett. 9 17 (2018) 5002-5008. doi:10.1021/acs.jpclett.8b02117.

[43] V. Amendola and M. Meneghetti, Laser ablation synthesis in solution and size manipulation of noble metal nanoparticles, Phys. Chem. Chem. Phys. 11(20) (2009) 3805–3821. doi: 10.1039/b900654k.

[44] F. Giammanco, E. Giorgetti, P. Marsili, and A. Giusti, Experimental and Theoretical Analysis of Photofragmentation of Au Nanoparticles by Picosecond Laser Radiation, J. Phys. Chem. C 114(8) (2010) 3354–3363. doi: 10.1021/jp908964t.

[45] R.G. Nikov, A.S. Nikolov, P.A. Atanasov, Preparation of gold and silver nanoparticles by pulsed laser ablation of solid target in water, 16th International School on Quantum Electronics: Laser Physics and Application, Proc. of SPIE Vol. 7747. doi: 10.1117/12.881908.

[46] B. Balamurugan and T. Maruyama, Size-modified d bands and associated interband absorption of Ag nanoparticles, J. Appl. Phys. 102 (2007) 034306. doi: 10.1063/1.2767837.

[47] A. S. Nikolov, N. E. Stankova, D. B. Karashanova, N. N. Nedyalkov, E. L. Pavlov, L. A. Avramov, K. Koev, 2019, Patent Application, Incoming number № 112999/24.09.2019 in Patent Office of Republic of Bulgaria, "Method for production of ultra-fine monodisperse nanoparticles with laser pulses".

[48] V. Amendola, O.M. Bakr, F. Stellacci, A Study of the Surface Plasmon Resonance of Silver Nanoparticles by the Discrete Dipole Approximation Method: Effect of Shape, Size, Structure, and Assembly, Plasmonics 5, (2010) 85–97, doi:10.1007/s11468-009-9120-4.

[49] U. Kreibig and M. Vollmer, Optical Properties of Metal Clusters, Springer, Heidelberg, 1995.

[50] A. Takami, H. Kurita, S. Koda, Laser-Induced Size Reduction of Noble Metal Particles, J. Phys. Chem. B 103(8) (1999) 1226-1232. doi:10.1021/jp983503o.

[51] V. Amendola and M. Meneghetti, Size Evaluation of Gold Nanoparticles by UV−vis Spectroscopy, J. Phys. Chem. C 113 (11) (2009) 4277–4285. doi: 10.1021/jp8082425.

[52] S. Chernousova, M. Epple, Silver as antibacterial agent: ion, nanoparticle, and metal, Angew Chem Int Ed Engl. 52(6) (2013) 1636-1653. doi: 10.1002/anie.201205923.

[53] L. Duta, C. Ristoscu, G.E. Stan, M.A. Husanu, C. Besleaga, M.C. Chifiriuc, V. Lazar, C.Bleotu, F. Miculescu, N. Mihailescu, E. Axente, M. Badiceanu, D. Bociaga, I.N. Mihailescu , New bio-active, animicrobial and adherent coatings of nanostructured carbon double-reinforced with silver and silicon by Matrix-Assisted Pulsed Laser Evaporation for medical applications, Applied Surface Science, 441 (2018) 871-883. https://doi.org/10.1016/j.apsusc.2018.02.047.

[54] G. Socol, M. Socol, L. Sima, S. Petrescu, M. Enculescu, F. Sima, M. Miroiu, G. Popescu-Pelin, N. Stefan, R. Cristescu, C.N. Mihailescu, A. Stanculescu, C. Sutan, I.N. Mihailescu, Combinatorial pulsed laser deposition of Ag-containing calcium phosphate coatings, Digest Journal of Nanomaterials and Biostructures, 7(2) (2012) 563 – 576.

[55] N. Hachicho, P. Hoffmann, K. Ahlert, H.J. Heipieper, Effect of silver nanoparticles and silver ions on growth and adaptive response mechanisms of Pseudomonas putida mt-2, FEMS Microbiol Lett. 355(1) (2014) 71-77, doi:10.1111/1574-6968.12460.